\documentstyle[art10]{article}
\begin{document}
\title{Renormalization in 1-D Quantum Mechanics: contact interactions}

\author{Luis J BOYA, Alejandro RIVERO \thanks{Dep. F\'{\i}sica
Teorica, Fac de Ciencias Univ de Zaragoza,
 50009 Zaragoza, Spain} \thanks{email: rivero@sol.unizar.es}}

\date{DFTUZ 9413; November 1994}
\maketitle

\begin{abstract}
 We have commemorated the 20th anniversary of the Wilson-Kogut review \cite{wk}
by building a toy model of the W-K RG in one dimensional
Quantum Mechanics... With it, we
show (well, sort of) that the RG flow in the set of
1-dimensional finite range S matrices fullfilling $S_{-k}^\dagger=QS_kQ$
defines the known four parametric set
of zero-range interactions.
\end{abstract}
%ICP: 03.65, 02.30 ??

\section{Introduction}

It is presented here a gadget model of the Wilson-Kogut \cite{wk}
renormalization group implemented in a Quantum Mechanical problem, a bit
following the mood of \cite{rgupta}.  But our scheme is complex enough
to be a good introduction before to go to full QFT-oriented reviews,
as the recent one from Ball-Thorpe \cite{cern-th}.

Examples of the renormalization group in QM have been built using the
traditional beta function setup (by example, see \cite{peter-rolf}) and,
recently, the path integral formalism \cite{polonyi}. To get a
non trivial W-K flow we work with QM on $R^1$. This one-dimensional setup
is richer (and more complicated) than typical "tridimensional" problems
in $R^3/O(3)$, which are usually reduced to problems
in the one dimensional half line. By working with
the full real line we are forced to calculate
in matrix form, which make the problem more illuminating
in the long way. This can be seen, by example, by comparing
Newton \cite{newton} vs. Fadeev \cite{fadeev} solutions of
the 1-D inverse scattering problem.

Our scheme moves close
to the standard studies of contact interactions:
self-adjoint extensions \cite{alb}, series of hamiltonians \cite{carreau},
regularizations \cite{cristina-rolf,jackiw} etc. So it can
illuminate some recent
conflicts in the literature, such as the status of
the controversial $\delta'$ interaction (which, btw, would be scale-invariant
in one dimension or at least to present characteristics close to
the $1/x^2$ studies from \cite{rgupta}).

This preprint represents work in course. Rigorization
of convergence issues in the perturbative analisis is in process.
Effort has been done to implement major features of the renormalization
group, but some interesting points, as C-functions
or correlation lengths are not implemented yet.

Plan of this paper is as follows: In section 1 we make some
introductory remarks and the plan of the paper is presented. Section
2 defines the interaction we are going to study and sketch some
needed formulae. In section 3 the
wilson-kogut RG is built and fixed points are calculated. Section 4
show some examples of trajectories got directly from
known solutions, for comparation.  Section 5 completes the topological
analysis of the RG flow calculating the stable and unstable directions
at fixed points. Sections 6 and 7 sketch some examples
showing how the mechanics of
of regularizated potentials and renormalized couplings. Both sections
are mainly didactical and only needed points are detailed.
We conclude in section 8 with some specific remarks about contact
interactions.

\section{The cut-off interaction}

\marginpar{We can relax this condition to be only "with range less than a"}
A localized interaction with cut-off $a$ will correspond to an interaction
which
is free out of the interval $(-a,a)$, but can have any form  in the
interior of this interval. So we work only with data external to $(-a,a)$.
Equivalently, it can be said that the cutoff "hides" or
"averages" any characteristic
of the interaction in distances lower than the cutoff, see figure 1.

We can characterize such interactions either by their
scattering matrix,
%vaya por dios, la ponemos al reves que newton.
\begin{equation}
%\label{callate}
S_k=\pmatrix{T^r & R^l \cr R^r & T^l}
\end{equation}
or by some matrix specifying the boundary conditions in $-a,a$. An useful
one, given its dimensional and scaling properties \cite{carreau}, is
$M \equiv \pmatrix{\alpha + \rho & -\rho e^{i\theta} \cr
   -\rho e^{-i\theta} & \beta + \rho}$
\begin{equation}
\pmatrix{-\psi'(-a) \cr \psi'(a)} =
             M^a_k \pmatrix{\psi(-a) \cr \psi(a)}
\end{equation}
Where the parameters in $M$ are reals, but can become indeterminates
or infinity for some interactions. In such case, we could use
other formulations \cite{seba,kurasov.delta}, closer to the standard
formalism of self adjoint extensions.

Note that, in principle -and forgetting some of inverse scattering theory-,
 different hamiltonians could be localized in the same
interval with equal conditions at the boundary; thus the $a$-cutoff
in some sense hides data about the interaction
to distance less than $2 a$.

The interaction being free out of this interval, the asymptotic
solution of the Schroedinger equation must remain valid over all
this zone $\{R-(-a,a)\}$.
Thus we can use the explicit
definition of the $S$-Matrix to connect the boundary conditions at both
sides of the interval. As we will need it in our examples, lets
skectch the formulae. For each eigenvalue $k$,
we can chose two independent solutions $u_1,u_2$ of the Schroedinger
 eq. fulfilling:
\begin{eqnarray}
u_1(x)= & e^{i k x} T^l & x>a \\
       & e^{i k x} + e^{-ikx} R^l & x<-a \\
u_2(x)= & e^{-ikx} + e^{ikx} R^r & x>a \\
       & e^{-ikx} T^r & x<-a
\end{eqnarray}
and evaluate them at $-a$ and $a$ to solve for the matrix $M^a_k$;
and reciprocally for $S_k$. We get the following relationships.
\begin{equation}
M_k=  -i k
 \pmatrix{
 { 1 - e^{2iak} R^l +  e^{2iak} R^r -  e^{4iak} R^l R^r + e^{4iak} T^l T^r
\over 1 + e^{2iak} R^l +  e^{2iak} R^r +  e^{4iak} R^l R^r - e^{4iak} T^l T^r}
&
 {2 T^r
\over 1 + e^{2iak} R^l +  e^{2iak} R^r +  e^{4iak} R^l R^r - e^{4iak} T^l T^r}
\cr {2 T^l
\over 1 + e^{2iak} R^l +  e^{2iak} R^r +  e^{4iak} R^l R^r - e^{4iak} T^l T^r}
&
{ 1 + e^{2iak} R^l -  e^{2iak} R^r -  e^{4iak} R^l R^r + e^{4iak} T^l T^r
\over 1 + e^{2iak} R^l +  e^{2iak} R^r +  e^{4iak} R^l R^r - e^{4iak} T^l T^r}}
\end{equation}
 \begin{equation}
S_k=- e^{-2 i a k} \pmatrix{
%T^r
{2 e^{i \theta} ik \rho
\over \alpha\beta-i\alpha k-i\beta k-k^2+\alpha\rho +\beta\rho -2 i k\rho} &
%R^l
{\alpha \beta-i\alpha k+i\beta k+ k^2+ \alpha \rho +\beta \rho
\over \alpha\beta-i\alpha k-i\beta k-k^2+\alpha\rho +\beta\rho -2 i k\rho} \cr
%R^r
{\alpha \beta+i\alpha k-i\beta k+ k^2+ \alpha \rho +\beta \rho
\over \alpha\beta-i\alpha k-i\beta k-k^2+\alpha\rho +\beta\rho -2 i k\rho}&
%T^l
{2 e^{-i \theta } ik \rho
\over \alpha\beta-i\alpha k-i\beta k-k^2+\alpha\rho +\beta\rho -2 i k\rho}}
\end{equation}

There are no problem going from one description to the other, as
here the only role of both $S_k$ and $M_k$ here is to
select a pair of eigenfunctions.
\footnote{BTW, We can check that making $a \to 0$ and requesting $S_k$ to be
unitary and complete,
the set of admisible solutions coincide with the result got
in \cite{carreau} , as $S_k$ unitary iff $M_k$ hermitian, and
$H^{in}=H^{out}=L^2(R)$ would imply $M_k$=cte when   $a = 0$.}

To be fully "Wilson-Kogut compliant" and draw the renormalization
flow in the space of fixed cut-off theories, it is need to work with
the adimensional matrices $\tilde S_{\tilde{k}},\tilde M_{\tilde{k}}$:
\begin{eqnarray}
%    \tilde S_{\tilde{k}} \equiv S_{\tilde{k}/a} &&
%    \tilde M_{\tilde{k}} \equiv a M_{\tilde{k}/a} \\
S_k \equiv \tilde S_{a k} &&
M_k \equiv {1 \over a} \tilde M_{a k}
\end{eqnarray}
\def\tk{{\tilde{k}}}
In this form, the relationship becomes:
\begin{equation}
\tilde M_\tk = -i \tk  \pmatrix{
 { 1 - e^{2i\tk} R^l +  e^{2i\tk} R^r - e^{4i\tk} R^l R^r + e^{4i\tk} T^l T^r
\over 1+e^{2i\tk}R^l+e^{2i\tk}R^r+e^{4i\tk}R^l R^r- e^{4i\tk} T^l T^r} &
{ 2 T^r
\over 1+e^{2i\tk}R^l+e^{2i\tk}R^r+e^{4i\tk}R^l R^r- e^{4i\tk} T^l T^r} \cr {2
T^l
\over 1+e^{2i\tk}R^l+e^{2i\tk}R^r+e^{4i\tk}R^l R^r- e^{4i\tk} T^l T^r} &
 {1 + e^{2i\tk} R^l -  e^{2i\tk} R^r - e^{4i\tk} R^l R^r + e^{4i\tk} T^l T^r
\over 1+e^{2i\tk}R^l+e^{2i\tk}R^r+e^{4i\tk}R^l R^r- e^{4i\tk} T^l T^r} }
\end{equation}
\begin{equation}
\tilde S_\tk = - e^{-2 i \tk}
\pmatrix{
%T^r
{2 e^{i \theta} i\tk \rho
\over \alpha\beta-i\alpha\tk-i\beta\tk-\tk^2+\alpha\rho+\beta\rho-2i\tk\rho} &
%R^l
{\alpha \beta-i\alpha \tk+i\beta \tk+ \tk^2+\alpha \rho +\beta \rho
\over \alpha\beta-i\alpha\tk-i\beta\tk-\tk^2+\alpha\rho+\beta\rho-2i\tk\rho}\cr
%R^r
{\alpha \beta+i\alpha \tk-i\beta \tk+ \tk^2+ \alpha \rho +\beta \rho
\over \alpha\beta-i\alpha\tk-i\beta\tk-\tk^2+\alpha\rho+\beta\rho-2i\tk\rho} &
%T^l
{2 e^{-i \theta } i\tk \rho
\over \alpha\beta-i\alpha\tk-i\beta\tk-\tk^2+\alpha\rho+\beta\rho-2i\tk\rho} }
\end{equation}
where now the ($k$-dependent) matrix terms $\alpha \beta \rho \theta$ refer
to the adim matrix $\tilde M_k$.

\section{The RG transformation, a la Wilson Kogut. Fixed Points}

As usual, we take the space $\{\tilde S\}$ of all the $a_0$-cutoff
interactions,
in dimensionless form. Each interaction can be given by a unitary $S(k)$,
which by standard scattering theory (see e.g. \cite{newton}) will
fulfill
\begin{equation}
\label{conjug}
S_{-k}^\dagger=QS_kQ
\end{equation}
where $Q=\pmatrix{0 & 1\cr 1 &0}$ and $\dagger$ is the hermitian conjugate.
We could in addition restrict ourselves to interactions invariant under
time reversal. In such case, we would add the condition $S_{-k}=
S_k^\star$, which implies $T^r=T^l$ and the know reciprocity
theorem
\begin{equation}
\label{recipr}
\hat S_k=QS_kQ
\end{equation}

Now, we define that two theories are in the same
line of the Renormalization Group flow if there are a pair of
scales $\{a,e^t a\}$ such that when we apply them to its respective theories,
we get the same physics (see figure 2), ie the same $S$-matrix in physical
dimensions.
Equivalently, $\tilde S^t_{\tilde{k}}$ will be the result of applying a RG
transformation
to $\tilde S_{\tilde{k}}$ iff
   $\tilde S_{a k} = \tilde S^t_{a e^t k}$,
this is,
  \begin{equation}
\label{rg.transf}
\tilde S^t_{\tilde{k}}=T^t[\tilde S_{\tilde{k}}] = \tilde S_{e^{-t}\tilde{k}}
\end{equation}

So the fixed points will be constant
$\tilde S$ matrices. This is, the subset of $U(2)$ fulfilling
property (\ref{conjug}), namely:
\begin{equation}
\{I_{\theta,\phi}\equiv \pmatrix{e^{i\theta}\cos\phi & -\sin\phi \cr
  \sin\phi & e^{-i\theta}\cos\phi}\} \cup \{\pm Q\}
\end{equation}

If we want to study only T-invariant potentials, we must add condition
(\ref{recipr}), and the set of fixed points reduces further to
\begin{equation}
\{\pmatrix{\cos \phi &-\sin \phi \cr \sin \phi &\cos \phi} \}  \cup \{ \pm Q \}
\end{equation}
Where the continuous circle of fixed points can be interpreted
corresponding to Kurasov $\delta'$ \cite{kurasov.delta}.
The rest of contact interactions is not scale-invariant, so we need
to study the flow near the fixed points to find them, as relevant
parameters.

Note that there are aditional asumptions on the analytic properties of
$S$, but we dont need to impose them to determine the fixed points, so
they will be commented when needed in section \ref{stability}.
At this point, note simply that the caracterization of space of interactions
is supossed to  be restricted to potentials with range smaller that the
cut-off.

\section{The RG flow, as seen from the QM solution}

Before entering in perturbative theory, it is good to give an
idea about what results we expect, to more easily follow the
argument. There are a four-parametric family of self-adjoint
extensions to the free hamiltonian in $R-\{0\}$. In this section
we calculate some subfamilies
of scattering matrices for known extensions and show its
form near a fixed point.

Following the standard theory(\cite{alb,seba,carreau}), lets take
contact-interaction given by the constant
matrix at a cutoff $0$.
\begin{equation}
M^{0}_k=\pmatrix{\alpha + \rho & -\rho e^{i\theta} \cr
   -\rho e^{-i\theta} & \beta + \rho}=cte.
\end{equation}
Its scattering matrix is:
\begin{equation}
S_k=- \pmatrix{
%T^r
{2 e^{i \theta} ik \rho
\over \alpha\beta-i\alpha k-i\beta k-k^2+\alpha\rho +\beta\rho -2 i k\rho} &
%R^l
{\alpha \beta-i\alpha k+i\beta k+ k^2+ \alpha \rho +\beta \rho
\over \alpha\beta-i\alpha k-i\beta k-k^2+\alpha\rho +\beta\rho -2 i k\rho}
\cr
%R^r
{\alpha \beta+i\alpha k-i\beta k+ k^2+ \alpha \rho +\beta \rho
\over \alpha\beta-i\alpha k-i\beta k-k^2+\alpha\rho +\beta\rho -2 i k\rho}&
%T^l
{2 e^{-i \theta } ik \rho
\over \alpha\beta-i\alpha k-i\beta k-k^2+\alpha\rho +\beta\rho -2 i k\rho}}
\end{equation}
which, using the lenght $a$ to remove dimensions, corresponds to a line
\def\tk{{\tilde{k}\over a}}
\def\pk{{(\tk)}}
\begin{equation}
\label{ren.line}
\tilde S_{\tilde{k},a} =
-\pmatrix{
%T^r
{2 e^{i \theta} i\tk \rho
\over \alpha\beta-i\alpha\tk-i\beta\tk-(\tk)^2+\alpha\rho+\beta\rho-2i\tk\rho}
&
%R^l
{\alpha \beta-i\alpha \tk+i\beta \tk+ (\tk)^2+\alpha \rho +\beta \rho
\over
\alpha\beta-i\alpha\tk-i\beta\tk-(\tk)^2+\alpha\rho+\beta\rho-2i\tk\rho}\cr
%R^r
{\alpha \beta+i\alpha \tk-i\beta \tk+ (\tk)^2+ \alpha \rho +\beta \rho
\over \alpha\beta-i\alpha\tk-i\beta\tk-(\tk)^2+\alpha\rho+\beta\rho-2i\tk\rho}
&
%T^l
{2 e^{-i \theta } i\tk \rho
\over \alpha\beta-i\alpha\tk-i\beta\tk-(\tk)^2+\alpha\rho+\beta\rho-2i\tk\rho}
}
\end{equation}
of renormalized interactions. By construction, the RG transformation
(\ref{rg.trans})
can be compensated by a change in the "spacing" (or cutoff) $a$.

As explained in figure 3, we expect solution lines  to be end-pointed by
fixed points. Specifically, we see that:

a) For $\rho=0; \alpha,\beta$ finite, which correspond to
two separate half-lines, the RG flow goes from
$
\tilde S_{\tilde{k}}= Q  %\pmatrix{0 & 1\cr 1& 0}
\mbox{\ to\ }
\tilde S_{\tilde{k}}=-Q   %\pmatrix{0 & -1\cr -1& 0}
$
We get this result in general for any  $\rho, \alpha,\beta$ finite
and different of zero.

b) For $\rho$ infinite, $\alpha,\beta$ finite, which for $\theta=0$
is the traditional $\delta$-interaction,
we get the flow going
\begin{equation}
\mbox{from \ }\tilde S_{\tilde{k}}=
\pmatrix{e^{i\theta} & 0\cr 0& e^{-i\theta}} \mbox{\ to\ }
\tilde S_{\tilde{k}}=-Q % \pmatrix{0 & -1\cr -1& 0}
\end{equation}
In particular, we  see that the fixed point governing the $\delta$
is the Identity.
\marginpar{/Identity/... but would we say the trivial fixed point?}

c) For $\rho$ finite but $\alpha=\beta=0$ (which when $\theta=0$ is the
so-called (by \cite{alb,carreau}) $\delta'$-interaction) the flow travels
along
\begin{equation}
\tilde S_{\tilde{k}}=Q % \pmatrix{0 & 1\cr 1& 0}
 \to
\tilde S_{\tilde{k}}=\pmatrix{e^{i\theta} & 0\cr 0& e^{-i\theta}}
\end{equation}

So all the interactions of this kind are to be governed either by
transparent interactions or by the totally reflective one. These
observations are summarized in figure 4.

It is instructive to look the interactions in the form $S=S^0+\Delta S$ near
a fixed point (around an endpoint, if we prefer to
ignore RG terminology in this section). We get for "+Q"
\begin{eqnarray}
\Delta S_a&&=\tilde S_{\tilde{k},a}-\pmatrix{0 & 1\cr 1& 0}= \\
\nonumber
&&=-\pmatrix{
%T^r
{2 e^{i \theta} i\tk \rho
\over \alpha\beta-i\alpha\tk-i\beta\tk-\pk^2+\alpha\rho+\beta\rho-2i\tk\rho} &
%R^l
2 {\alpha \beta-i\alpha \tk+\alpha \rho +\beta \rho- i\tk\rho
\over \alpha\beta-i\alpha\tk-i\beta\tk-\pk^2+\alpha\rho+\beta\rho-2i\tk\rho}\cr
%R^r
2{\alpha \beta-i\beta \tk+ \alpha \rho +\beta \rho-i\tk\rho
\over \alpha\beta-i\alpha\tk-i\beta\tk-\pk^2+\alpha\rho+\beta\rho-2i\tk\rho} &
%T^l
{2 e^{-i \theta } i\tk \rho
\over \alpha\beta-i\alpha\tk-i\beta\tk-\pk^2+\alpha\rho+\beta\rho-2i\tk\rho} }
\end{eqnarray}
for $-Q$:
\begin{eqnarray}
\Delta S_a&&=\tilde S_{\tilde{k},a}-\pmatrix{0 & -1\cr -1& 0}= \\
\nonumber
&&=-\pmatrix{%T^r
{2 e^{i \theta} i\tk \rho
\over \alpha\beta-i\alpha\tk-i\beta\tk-\pk^2+\alpha\rho+\beta\rho-2i\tk\rho} &
%R^l
2{+i\beta \tk+ \pk^2 + i\tk\rho
\over \alpha\beta-i\alpha\tk-i\beta\tk-\pk^2+\alpha\rho+\beta\rho-2i\tk\rho}\cr
%R^r
2{+i\alpha \tk+ \pk^2+  i\tk\rho
\over \alpha\beta-i\alpha\tk-i\beta\tk-\pk^2+\alpha\rho+\beta\rho-2i\tk\rho} &
%T^l
{2 e^{-i \theta } i\tk \rho
\over \alpha\beta-i\alpha\tk-i\beta\tk-\pk^2+\alpha\rho+\beta\rho-2i\tk\rho} }
\end{eqnarray}
and for each $I_{\theta,0}$:
\begin{eqnarray}
\Delta S_a&&=\tilde S_{\tilde{k},a}-\pmatrix{e^{i \theta} &0\cr 0&  e^{-i\theta
}}= \\
\nonumber
&&=-\pmatrix{%T^r
{\alpha\beta-i\alpha\tk-i\beta\tk-\pk^2+\alpha\rho+\beta\rho
\over \alpha\beta-i\alpha\tk-i\beta\tk-\pk^2+\alpha\rho+\beta\rho-2i\tk\rho}
e^{i\theta} &
%R^l
{\alpha \beta-i\alpha \tk+i\beta \tk+ \pk^2+\alpha \rho +\beta \rho
\over \alpha\beta-i\alpha\tk-i\beta\tk-\pk^2+\alpha\rho+\beta\rho-2i\tk\rho}\cr
%R^r
{\alpha \beta+i\alpha \tk-i\beta \tk+ \pk^2+ \alpha \rho +\beta \rho
\over \alpha\beta-i\alpha\tk-i\beta\tk-\pk^2+\alpha\rho+\beta\rho-2i\tk\rho} &
%T^l
{ \alpha\beta-i\alpha\tk-i\beta\tk-\pk^2+\alpha\rho+\beta\rho
\over \alpha\beta-i\alpha\tk-i\beta\pk-\tk^2+\alpha\rho+\beta\rho-2i\tk\rho}
e^{-i\theta} }
\end{eqnarray}

\marginpar{(Notar que $\Delta S$ tambien va con determinante uno)}

With this, we can see outgoing and ingoing trajectories near a fixed point:

-Lines starting from $+Q$  with $\rho=0$
\begin{eqnarray}
\Delta S_a
&&=-\pmatrix{
%T^r
0 &
%R^l
2  {\alpha \beta-i\alpha \tk
\over \alpha\beta-i\alpha\tk-i\beta\tk-\pk^2}\cr
%R^r
2{\alpha \beta-i\beta \tk
\over \alpha\beta-i\alpha\tk-i\beta\tk-\pk^2} &
%T^l
0 }
=-\pmatrix{
0 &
2  {\alpha \over \alpha-i\tk}\cr
2{\beta \over \beta-i\tk} &
0 }= \\
&&=-2 \pmatrix{ 0 & {1 \over 1-{i\over \alpha} \tk}\cr
                {1 \over 1-{i\over \beta} \tk} &  0 }
\approx_{a<<1} \pmatrix{ 0 & -i {2 \alpha a \over {\tilde k}} \cr
                -i {2 \beta a \over {\tilde k} } &  0 }
\end{eqnarray}

-Starting from "+Q" with $\alpha=\beta=0$
\begin{eqnarray}
\Delta S_a&&=\tilde S_{\tilde{k},a}-\pmatrix{0 & 1\cr 1& 0}=
=-\pmatrix{
%T^r
{2 e^{i \theta} i \rho
\over -\tk-2i\rho} &
%R^l
2 {- i\rho
\over -\tk-2i\rho}\cr
%R^r
2{-i\rho
\over -\tk-2i\rho} &
%T^l
{2 e^{-i \theta } i \rho
\over -\tk-2i\rho} }=\\
&&=-\pmatrix{  -e^{i \theta} & 1 \cr 1 & -e^{-i \theta}}
        {1 \over 1-{i\over 2 \rho} \tk}
\approx -i \pmatrix{  -e^{i \theta} & 1 \cr 1 & -e^{-i \theta}}
       {2 \rho a \over \tilde k}
\end{eqnarray}

-From $I_{\theta,0}$ with $\rho \to \infty$:
\begin{eqnarray}
\Delta S_a=&&\tilde S_{\tilde{k},a}-\pmatrix{e^{i \theta} &
                0\cr 0&  e^{-i\theta }}
=-\pmatrix{%T^r
{+\alpha+\beta \over \alpha+\beta-2i\tk} &
%R^l
{+\alpha  +\beta  \over \alpha+\beta-2i\tk}\cr
%R^r
{+ \alpha  +\beta  \over \alpha+\beta-2i\tk} &
%T^l
{ +\alpha+\beta \over \alpha+\beta-2i\tk} }
=\\
&&-\pmatrix{ 1 & 1 \cr 1 & 1}
        {1 \over 1-{2 i\over (\alpha+\beta)} \tk}
\approx -\pmatrix{ i & i \cr i & i}
        {2  (\alpha+\beta) a \over \tilde k}
\end{eqnarray}

Now, for the incoming lines we define $\b{a} \equiv 1/a$, so $\b{a}<<1$
and we get:

-Lines incoming to $I_\theta$, $\alpha=\beta=0$.
\def\tk{{\b{a} \tilde{k}}}
\begin{eqnarray}
\Delta S_{\b{a}}&&=
-\pmatrix{%T^r
{-\tk \over -\tk-2i\rho} &
%R^l
{+ \tk \over  -\tk-2i\rho}\cr
%R^r
{+ \tk \over  -\tk-2i\rho} &
%T^l
{ -\tk \over  -\tk-2i\rho} }
=\\
&&=-\pmatrix{ 1 & -1 \cr -1 & 1}
        {1 \over 1-{2 \rho \over i \tk}}
\approx
 - \pmatrix{ -i & i \cr i & -i}
        {\tk \over 2 \rho}
\end{eqnarray}

-Arriving to $-Q$: (a) with $\rho=0$
\begin{eqnarray}
\Delta S_{\b{a}}&&=
-\pmatrix{%T^r
0 &
%R^l
2{+i\beta \tk+ \tk^2
\over \alpha\beta-i\alpha\tk-i\beta\tk-\tk^2}\cr
%R^r
2{+i\alpha \tk+ \tk^2
\over \alpha\beta-i\alpha\tk-i\beta\tk-\tk^2} &
%T^l
0 }
= \\
&&= -\pmatrix{%T^r
0 &
%R^l
2{+i\tk \over \alpha-i\tk}\cr
%R^r
2{+i\tk \over \beta-i\tk} &
%T^l
0 }
=-2 \pmatrix{ 0 & {1 \over 1-{\alpha \over i \tk}}\cr
                {1 \over 1-{\beta \over i \tk}} &  0 }
\\
&& \approx - \pmatrix{ 0 & -{2\over \alpha} i \cr
              -{2\over \beta}i & 0} \tk
\end{eqnarray}
and from lines type (b):
\begin{eqnarray}
\Delta S_{\b{a}}&&=
=-\pmatrix{%T^r
{2 e^{i \theta} i\tk \over \alpha+\beta-2i\tk} &
%R^l
2{ i\tk \over \alpha+\beta-2i\tk}\cr
%R^r
2{  i\tk \over \alpha+\beta-2i\tk} &
%T^l
{2 e^{-i \theta } i\tk \over \alpha+\beta-2i\tk } }  = \\
&&= -\pmatrix{  -e^{i \theta} & -1 \cr -1 & -e^{-i \theta}}
        {1 \over 1-{(\alpha+\beta) \over 2 i \tk} }
\approx - i \pmatrix{e^{i \theta} & 1 \cr 1 & e^{-i \theta}}
         {2\over \alpha + \beta} \tk
\end{eqnarray}

(Note that the approximations here are given in a non rigurous
way, simply to have a reference for the next section)

\marginpar{...or to give dimensions to $\tilde k$}
It's worth to note that the coupling constants appear
clearly related to the
(dimensional) constant we used to remove dimensions of $k$.
Compare e.g. with
 \cite{cristina-rolf,peter-rolf}.

\section{Stability. Relevant et irrelevant directions}

\label{stability}
\marginpar{getting ourselves out of the forest...}

Now, we need to develop a perturbation theory\footnote{The model is
simple enough to be exactly solved, but we consider more didactical to
remain close to wilson-kogut papers} around the fixed point directly
in the S-matrix formalism. In some neighbourhood of the identity,  where
the exponential map is one ot one, we can use the generators ${L_i}$ of
the $U(2)$ group, and write the perturbed
system as
\begin{equation}
S^{\vec a}= S_0 e^{{\vec a}(k) \cdot {\vec L}}
\end{equation}
with ${\vec a}(k) \in R^4$.For $\int ||{\vec a_k}|| dk$  small
\footnote{for simplicity, we are going to be a bit loose with this
condition}, we can put it as:
\begin{equation}
S^{\vec a}-S_0 \approx {\vec a}(k) \cdot (S_0 {\vec L})
\end{equation}

 The generators of U(2) are
\begin{equation}
L=\{
\pmatrix{ -i &  0 \cr 0& -i },
\pmatrix{ 0& i\cr i& 0},
\pmatrix{ 0& 1\cr -1& 0},
\pmatrix{ i& 0\cr 0& -i} \}
\end{equation}
(remember that $-i L= \{ I, \sigma_x,\sigma_y,\sigma_z \}$
So, around +Q we get
\begin{equation}
S^{+Q}_0 L = \{
\pmatrix{0 & -i \cr -i& 0},
\pmatrix{i & 0\cr 0& i},
\pmatrix{-1 & 0\cr 0& 1},
\pmatrix{0 & -i \cr i& 0} \}
\end{equation}
And for the other fixed points we have
\begin{equation}
S^{-Q}_0 L = -S^{+Q}_0 L,
S^{I_{0,0}} L= L,
S^{I_{\theta,\phi}}_0 L = ...
\end{equation}

As our space of interactions impose restrictions to the
admisible $S_k$, we get restrictions on ${\vec a}(k)$  depending
on the fixed point. Imposing condition (\ref{conjug})
around $I_0$ we get
\begin{eqnarray}
a_0(-k)=-a_0(k) \\
a_1(-k)=-a_1(k)  \\
a_2(-k)= a_2(k)  \\
a_3(-k)= a_3(k)
\end{eqnarray}
Of course, the additional asumption (\ref{recipr}) of
reciprocity imposes $a_3(k)=0$.

Now, we put the RG transformation in differential form.

\begin{equation}
T^{\delta t} S_k= S_{e^{-\delta t} k} \approx S_{k} + (O_k S_k) \delta t
\end{equation}
where
\begin{eqnarray}
O_k S_k = {\partial S_{e^{-t} k} \over \partial t} |_{t=0}=
        - S'( k )  k
\end{eqnarray}

We are looking for vectors ${\vec a}(k)$ such that
\begin{equation}
 T^{\delta t} S^{{\vec a}(k)} \approx S^{\lambda {\vec a}(k)}
\end{equation}
which to first-order amounts to
\begin{equation}
 S^0 + {\vec a}(k) S^0\vec L -  {\vec a}'(k) S^0\vec L k \delta t
\approx S^0 + \lambda {\vec a}(k) S^0\vec L
\end{equation}

So we have an eigenvalue equation
\begin{equation}
 - k {\vec a}'(k) \vec{S^0L} {\delta t}=(\lambda -1){\vec a}(k) \vec{S^0L}
\end{equation}
which solves to
\begin{eqnarray}
(\lambda -1)  = n  \delta t\\
{\vec a}(k)= k^{-n} {\vec a_0}
\end{eqnarray}

As an example, lets calculate the flow near $I_0$ with some detail.
It could seem that the valid ${\vec a}(k)$ would be
\begin{eqnarray}
\{ 0 ,0, 1,0 \} & marginal& \lambda=1\\
\{ 0 ,0, 0,1 \} &  marginal&\\
\{ k ,0, 0 ,0 \} & irrelevant & \lambda=1-\delta t\\
\{ 0 ,k, 0 ,0 \} &  irrelevant& \\
\{ 1/k ,0, 0 ,0 \} & relevant& \lambda=1+\delta t\\
\{ 0,1/k, 0 ,0 \} &  relevant&
\end{eqnarray}

Now, finite range condition when required in even potentials
implies \footnote{Well, I dont know of any rigourous proof
at this moment...} for the fase shifts at small $k$
\begin{eqnarray}
\mbox{even wave}:&&\tan \delta_0 \approx 1/k \\
\mbox{odd wave}:&&\tan \delta_1 \approx k
\end{eqnarray}
So, this condition for our space of interactions
rules out the combinations
\begin{eqnarray}
\{ -k , k, 0 ,0 \}  \\
\{ -1/k , -1/k , 0 ,0 \}
\end{eqnarray}
as well as higher orders in $k$, and let us
with four directions,
\begin{eqnarray}
\label{dir.kurasov} \{ 0 ,0, 1,0 \} & marginal& \lambda=1\\
\label{dir.no.t} \{ 0 ,0, 0,1 \} &  marginal& \lambda=1\\
\label{dir.albeverio} \{ k ,k, 0 ,0 \} & irrelevant & \lambda<1\\
\label{dir.delta} \{ 1/k ,-1/k, 0 ,0 \} & relevant& \lambda>1
\end{eqnarray}
which coincides with the known result, derived in the
previous section. Here, the marginal direction (\ref{dir.kurasov}) can
be asociated with the circle of kurasov' $\delta^\prime$ family;
the relevant one (\ref{dir.delta}) is the usual $\delta$, and the
irrelevant parameter one (\ref{dir.albeverio}) can be seen
 as coming from the fixed
point $+Q$, then producing the line corresponding to Albeverio
et al. so-called $\delta^\prime$ (which happens to be not
scale-invariant). Finally, lets note that direction (\ref{dir.no.t})
produces a family of scale invariant interactions which haven't
time-reversal symmetry.

\section{Regularizations and its flow}

Given a series of cutoff interactions
the RG mechanism, as explained in figure 3, let us to obtain.
a renormalized interaction at a given scale $a_0$,

Lets see, as first example, the series of effective pseudopotentials
$\{V_a\}$ proposed by Carreau \cite{carreau}. Each $V_a$ is zero
out of the interval $(-a,a)$, and the M matrix at $\{-a,a\}$
always the same and independent of $k$. In such case, we get
a series $\{\tilde M^a \equiv a M_0\}$ in the space of
dimensionless cutoff interactions. The $\tilde S$ matrix is:
\def\tk{{\tilde{k} \over a}}
 \begin{equation}
\tilde S_{\tilde{k}} = - e^{-2 i \tilde{k}}
\pmatrix{
%T^r
{2 e^{i \theta} i\tk \rho
\over \alpha\beta-i\alpha\tk-i\beta\tk-\pk^2+\alpha\rho+\beta\rho-2i\tk\rho} &
%R^l
{\alpha \beta-i\alpha \tk+i\beta \tk+ \pk^2+\alpha \rho +\beta \rho
\over \alpha\beta-i\alpha\tk-i\beta\tk-\pk^2+\alpha\rho+\beta\rho-2i\tk\rho}\cr
%R^r
{\alpha \beta+i\alpha \tk-i\beta \tk+ \pk^2+ \alpha \rho +\beta \rho
\over \alpha\beta-i\alpha\tk-i\beta\tk-\pk^2+\alpha\rho+\beta\rho-2i\tk\rho} &
%T^l
{2 e^{-i \theta } i\tk \rho
\over \alpha\beta-i\alpha\tk-i\beta\tk-\pk^2+\alpha\rho+\beta\rho-2i\tk\rho} }
\end{equation}
where now the parameters $\alpha,\beta,\rho,\theta$ are the
constants of the initial matrix $M$. The limit $a\to 0$ is
$\tilde S^0=  - e^{-2 i \tilde{k}} \tilde S^{fp}$, where $\tilde S^{fp}$ one
of the fixed points studied in section 5. Obviously the
RG transformation moves $S^0$ towards $S^{fp}$; so
when using the RG  we will move near $S^{fp}$
and the renormalized series will converge to a renormalized
interaction in the relevant line. To be concrete, we begin with
a cut-off $a_0$, and for each $a$ we recover the original scale
by applying $T^{\log (a_0/a)}$, thus getting:
\def\tk{{\tilde{k} \over a_0}}
\begin{equation}
(T \tilde S)_{\tilde{k}} = - e^{-2 i \tilde{a k \over a_0}}
\pmatrix{
%T^r
{2 e^{i \theta} i\tk \rho
\over \alpha\beta-i\alpha\tk-i\beta\tk-\tk^2+\alpha\rho+\beta\rho-2i\tk\rho} &
%R^l
{\alpha \beta-i\alpha \tk+i\beta \tk+ \tk^2+\alpha \rho +\beta \rho
\over \alpha\beta-i\alpha\tk-i\beta\tk-\tk^2+\alpha\rho+\beta\rho-2i\tk\rho}\cr
%R^r
{\alpha \beta+i\alpha \tk-i\beta \tk+ \tk^2+ \alpha \rho +\beta \rho
\over \alpha\beta-i\alpha\tk-i\beta\tk-\tk^2+\alpha\rho+\beta\rho-2i\tk\rho} &
%T^l
{2 e^{-i \theta } i\tk \rho
\over \alpha\beta-i\alpha\tk-i\beta\tk-\tk^2+\alpha\rho+\beta\rho-2i\tk\rho} }
\end{equation}
and we see that in the limit $a\to 0$ we recover (\ref{ren.line}), as expected.
%hipotesis: quizas todas las teorias en la superficie critica sean
% las de la forma (exponencial . matriz constante)

To go for a more complicated example, lets use the two-deltas regulator for
$\delta'$ interaction (this would be as a core-shell regularization),
$V={g \over 2a} (\delta(x+a)- \delta(x-a))$

the matching conditions are:
\begin{eqnarray}
{2ika \over g} ((A-1)e^{-ika}-(B-R^l) e^{ika})=
e^{-ika}+R^l e^{ika}= A e^{-ika} + B e^{ika}
\\
-{2ika \over g} (B e^{-ika} - (A-T^l) e^{ika}) =
T^l e^{ika}= B e^{-ika} + A e^{ika}
\end{eqnarray}
The scattering matrix is:
\begin{equation}
\label{S2d}
S_k^a=\pmatrix{
%T^r
{1 \over 1 - ({g \over 4ak})^2 (e^{4ika}-1) }&
%Rl
e^{-2ika}{ (e^{4ika}-1) (1-{ig \over 4ak}) \over {4ak \over ig} +
   {ig \over 4ak} (e^{4ika}-1) } \cr
%Rr
...&
%Tl
...\cr }
\end{equation}
which goes to $-Q$ as $a \to 0$.

In the adim space this stuff becomes:
\def\tk{{\tilde{k}}}
\begin{eqnarray}
\label{implicit}
{i 2 \tk \over g} ((A-1)e^{-i\tk}-(B-R^l) e^{i\tk})=
e^{-i\tk}+R^l e^{i\tk}= A e^{-i\tk} + B e^{i\tk}
\\
\nonumber
-{i2 \tk \over g} (B e^{-i\tk} - (A-T^l) e^{i\tk}) =
T^l e^{i\tk}= B e^{-i\tk} + A e^{i\tk}
\end{eqnarray}
\begin{equation}
\label{S2d.adim}
\tilde S_{\tilde k}^a=\pmatrix{
%T^r
{1 \over 1 - ({g \over 4\tk})^2 (e^{4i\tk}-1) }&
%Rl
{ (e^{2i\tk}-e^{-2i\tk}) (1-{ig \over 4\tk}) \over {4\tk \over ig} +
   {ig \over 4\tk} (e^{4i\tk}-1) } \cr
%Rr
...&
%Tl
...\cr }
\end{equation}
 and the limits are:
\begin{eqnarray}
%S_k^{a \to 0}=\pmatrix{0 & -1 \cr ...&   ...      }
%&& S_{k \to 0}^a=\pmatrix{0 & -1 \cr ... &0} \\
\tilde S_{\tilde k}^{a \to 0}= \tilde S_{\tilde k}^{a}
&& \tilde S_{{\tilde k}\to 0}^a=\pmatrix{0 & -1 \cr -1 &0}
\end{eqnarray}

This shows the qualitative diference between both
formalisms. In (\ref{S2d}) we simply take the
limit $a \to 0$ expecting in to be well-behaved (as it happens in this
simple case). In the RG approach, we first got the limit point, and then
we look for the fixed point atracting it.

Here, any limiting procedure will
carry us inexorably to the Dirichlet fixed point (see Seba).
If we want
to get a non trivial result, we need to implementent a
dependence for the coupling constant. This can be seen a la Tarrach
in the
implicit equations;  if $g(a) \to 0$ the eq (\ref{implicit}) has a
indetermination and we will need go to g'(a).

\section{Coupling constant renormalization}

Lets continue with the previous example. We ask for a $g(a)$ dependence
giving us a non trivial limit. The example is simple enough to
directly read the answer from (\ref{S2d}). Regretly the RG mechanism
in QM is too simple \cite{rgupta} and it is not possible
to get remarkable differences. Lets sketch the method anyway.

Equation (\ref{S2d.adim}) let us define a subset $\{\tilde S^a(g)\}$
of interactions
in the space $S$. We need to get series $\tilde S^a\equiv
S_{k/a}(g(a))$ such that the
limit point $a \to 0$ falls in the atracttion point
of a non trivial fixed point. Any $g(a)$ going to zero as $a \to 0$
makes the trick, falling directly in the fixed point $I_{0,0}$.
Furthermore, we want the corresponding
renormalized series $T^{-\log t(a)}\tilde S^a$ to have a
non trivial limit. This is enforced in the usual manner,
asking for no dependence of $a$ in the limit. This
is get by putting
\begin{equation}
t(a)= \alpha g^2(a)
\end{equation}
and then
\begin{equation}
\lim_{a\to 0} T^{-\log t(a)}\tilde S^{a}_{\tilde k}(g(a))=
%\pmatrix{
%T^r
%{1 \over 1 - ({1\over a_0} a {1\over 4\tk})^2 (e^{4i\tk}-1) }&
%Rl
%{ (e^{2i\tk}-e^{-2i\tk}) (1-{ig \over 4\tk}) \over {4\tk \over ig} +
%   {ig \over 4\tk} (e^{4i\tk}-1) } \cr
%Rr
%...&
%Tl
%...\cr }
\pmatrix{
%T^r
{1 \over 1 - ({1\over \alpha \tk})}&
%Rl
...  \cr
%Rr
...&
%Tl
...\cr }
\end{equation}
which is the S-matrix of the  $\delta$ interaction. Of course, if we
put $t(a)=a/a_0$, as given by the usual scaling, we get
\begin{equation}
g(a)= {1 \over a_0 \alpha} a^{1/2}
\end{equation}
So we have got an alternate derivation of the known result of
Seba \cite{seba}.

\marginpar{there are difference between $Lim_k Lim_a$
and $lim_a lim_k$. If both limits conmute, would we fall into the
scale-invariant interaction, the fixed point?
}

\section{Remarks}

We can always get a known regularization of the $\delta$ or
the $\delta '$ and look for the renormalized interaction. It  can
be got  partial
but important information simply taking the limit of the unrenormalized
series and asking which fixed point is reached when applying the RG to this
limit interaction. By example, lets note that even if the
nonrenormalized interaction falls in the atraction domain of
the two half-lines fixed point, $\tilde S=+Q$,
%(it happens with the
%regularization using 2 opossed $\delta$ functions at a distance 2 $a$,
%or even with linear aproximations to this)
 it is unlikely to reach any interaction in
the renormalized line if the series happens
to fall in the domain
of  $\tilde S=-Q$.

In particular, if we want to reach a limit of the kind of Albeverio et
al. "$\delta^\prime$", we will need series of interactions in the
attraction domain of the $I_{0,0}$, and with its limit in the
domain of $+Q$. Such properties seem to imply that any regularization
for this interaction would fullfill $S_{k\to 0} \to I$, which greatly
restricts the class of candidates.

%\section{Acknowledgements}

%We would like to thank J. Esteve and J. Casahorran by useful discussions
%and pointers.
 This job has been supported in part with funds provided
by CICYT (Spain). The
author must acknowledge grant AP9029093359.

%\section{Reference}

\bibliographystyle{plain}
%\bibliography{tesis/refs}

%\begin{thebibliography}{20}
%\end{thebibliography}
%\end{document}
%\bibitem{coleman} S. COLEMAN, ``Aspects of Symmetry'', Cambridge Univ. Press,
%%%1985
%\bibitem{simons} B. SIMON,
%Semiclassical analysis of low lying eigenvalues II: Tunneling,
% {\em Annals of Mathematics}, {\bf 120} (1984) 84

%captions: remove the \end document to print them.
% \enddocument

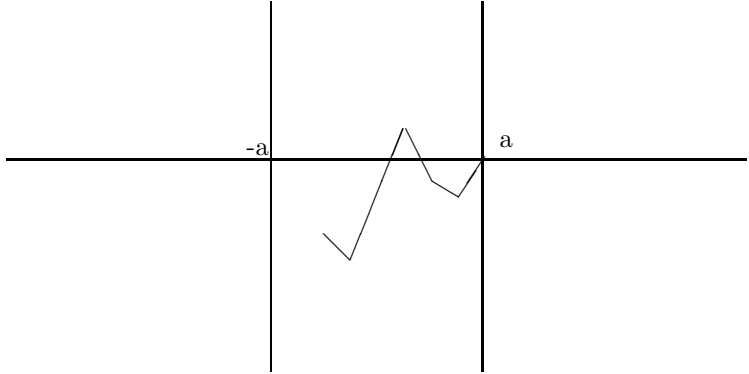
\begin{figure}
{\unitlength=1.000000pt
\begin{picture}(300.00,160.00)(300.00,0.00)
\put(450.00,62.00){\line(-1,1){10.00}}
\put(450.00,62.00){\line(2,5){20.00}}
\put(380.00,72.00){\line(0,1){0.00}}
\put(300.00,0.00){\oval(0.00,0.00)}
\put(340.00,20.00){\rule{0.00\unitlength}{0.00\unitlength}}
\put(500.00,20.00){\line(0,1){0.00}}
\put(500.00,160.00){\line(0,-1){140.00}}
\put(500.00,160.00){\line(0,1){0.00}}
\put(420.00,160.00){\line(0,-1){140.00}}
\put(320.00,100.00){\line(1,0){280.00}}
\put(509.00,107.00){\makebox(0.00,0.00){a}}
\put(415.00,104.00){\makebox(0.00,0.00){-a}}
%\put(425.00,100.00){\line(5,-13){10.00}}
\put(471.00,112.00){\line(1,-2){10.00}}
\put(481.00,92.00){\line(5,-3){10.00}}
\put(491.00,86.00){\line(2,3){10.00}}
\end{picture}}
\caption{Any information about the interaction
to distances shorter that the cutoff is hidden "under the cutoff"}
\label{uno}
\end{figure}

\begin{figure}
{\unitlength=1.000000pt
\begin{picture}(370.00,110.00)(60.00,0.00)
\put(390.00,110.00){\line(0,-1){110.00}}
\put(310.00,110.00){\line(0,-1){110.00}}
\put(360.00,80.00){\line(0,-1){30.00}}
\put(350.00,70.00){\line(1,1){10.00}}
\put(350.00,50.00){\line(0,1){20.00}}
\put(350.00,30.00){\line(0,1){20.00}}
\put(340.00,30.00){\line(1,0){10.00}}
\put(340.00,50.00){\line(0,-1){20.00}}
\put(280.00,50.00){\line(1,0){150.00}}
\put(200.00,50.00){\vector(1,0){50.00}}
\put(130.00,100.00){\line(0,-1){100.00}}
\put(90.00,100.00){\line(0,-1){100.00}}
\put(120.00,80.00){\line(0,-1){30.00}}
\put(110.00,70.00){\line(1,1){10.00}}
\put(110.00,30.00){\line(0,1){40.00}}
\put(100.00,30.00){\line(1,0){10.00}}
\put(100.00,50.00){\line(0,-1){20.00}}
\put(60.00,50.00){\line(1,0){110.00}}
\end{picture}}
\caption{The renormalization group transformation joins theories with
the same physics but different cut-off}
\label{dos}
\end{figure}
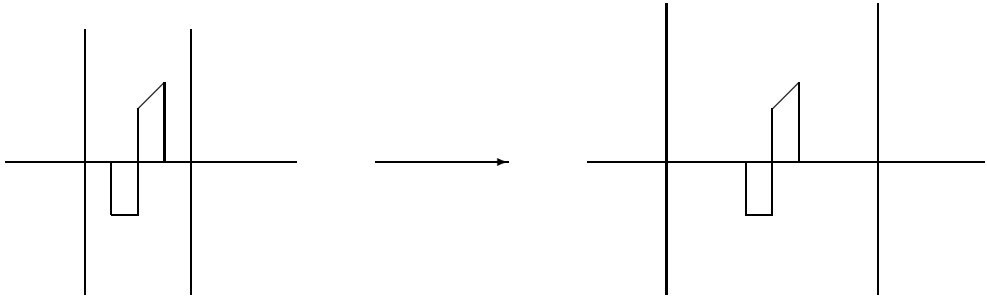

\begin{figure}
\caption{Renormalization group scheme}
Line A contains a series of unrenormalized theories with
decreasing cutoff. The renormalization group transformations let
us to map this to one series (line C) of theories with the same cutoff,
say $a_0$.
Such series has as limit a point in the line B of renormalized
interactions ($T^\infty\{\tilde S \}$). The flow corresponding to renormalized
interactions
is limited by fixed points (endpoints of B), but any other theory could
be driven out of the space of interactions when integrated back
with the renormalization group transformation (case D).
\label{tres}
\end{figure}

\begin{figure}
\caption{Expected topology of the RG flow (artist's view)}
The usual $\delta$ potential corresponds to the line from
$\pmatrix{1 & 0 \cr 0 & 1}$ to $\pmatrix{0 & -1 \cr -1 & 0}$ .
It is unlikely to reach the "$\delta '$ line" ($\alpha=\beta=0$) by
renormalizing interactions
in the domain of the $\pmatrix{0 & -1 \cr -1 & 0}$ fixed point, so
some regularizations will give us renormalized interactions in the
line of the $\delta$.
Note that this drawing is somehow a projection of the $\infty$-dim space,
and RG trajectories doesn't cross in reality.
\label{cuatro}
\end{figure}

\end{document}